\documentclass[a4paper,fleqn,usenatbib]{mnras}
\usepackage{graphicx}
\usepackage{amsmath}
\usepackage{amssymb}
\usepackage{times,txfonts}
\usepackage[T1]{fontenc}
\usepackage{ae,aecompl}
\usepackage{graphicx}

\title[Improved orbits and parallaxes]{Improved orbits and parallaxes for eight visual binaries with unrealistic previous masses using the Hipparcos parallax}

\author[J. A. Docobo et al.]{J. A. Docobo$^{1}$\thanks{E-mail: joseangel.docobo@usc.es (JAD), vakhtang.tamazian@usc.es (VST), malkov@inasan.ru (OYM), pedropablo.campo@usc.es (PPC), chulkov@inasan.ru (DAC)}, V. S. Tamazian$^{1}$, O. Yu. Malkov$^{2}$, P. P. Campo$^{1}$, D. A. Chulkov$^{2}$
\\
$^{1}$Observatorio Astron\'omico Ram\'on Mar\'{\i}a Aller, Universidade de Santiago de Compostela, Avenida das Ciencias s/n, 15782\\
Santiago de Compostela, Spain.\\
$^{2}$Institute of Astronomy of the Russian Academy of Sciences, 48 Pyatnitskaya Street, Moscow 119017, Russia. \\
}
\date{Accepted XXX. Received YYY; in original form ZZZ}
\pubyear{2016}
\begin{document}
\label{firstpage}
\pagerange{\pageref{firstpage}--\pageref{lastpage}}
\maketitle
\begin{abstract}

Improved orbits are presented for the visual binaries WDS 02366+1227, WDS 02434-6643, WDS 03244-1539, WDS 08507+1800, WDS 09128-6055, WDS 11532-1540, WDS 17375+2419, and WDS 22408-0333. The latest orbits for these binaries were demonstrating a great inconsistency between the systemic mass obtained through Kepler's Third Law and that calculated as a sum of their components' mass through standard mass-luminosity and mass-spectrum relationships. Our improvement allowed us to obtain consistent systemic masses for WDS 02434-6643 and WDS 09128-6055 without a need for changing the Hipparcos parallax. For the remaining 6 pairs, we suggest the use of their dynamical parallax as a reliable distance estimate unless more precise parallax is reported. Astrophysical and dynamical properties of individual objects are discussed.

\end{abstract}


\section{Introduction}
Determination of accurate orbits in visual binaries with known distances represents a reliable method to obtain their systemic mass using Kepler's Third Law. However, its direct application sometimes leads to unrealistic mass values largely inconsistent with the mass sum of individual components derived from empirical mass-luminosity and mass-spectrum calibrations. A careful inspection of their orbits and/or estimated distances is needed in order to clarify the reasons and nature of such inconsistent data. Essentially (but not only), these are poorly known (or even erroneous) distance determination, low quality (preliminary) orbital solution, and/or the previously unknown multiplicity of the components. In a recent paper of Tamazian et al. (2016) we have selected a small sample of 17 such binaries and have presented brief comments regarding how such consistency can be fixed either by improving the orbit or by refining the distance. Here we report a detailed analysis of new orbits from this sample and discuss their astrophysical and dynamical characteristics as well.

\section{New orbits}
The method of Docobo (1985; 2012) was applied in orbital calculations, using a whole set of available measurements for each star taken mainly from the SIDONIe database (Le Contel et al. 2001) and the INT4 Catalog (Hartkopf et al. 2001a). Washington Double Star (WDS) designation is used for identification purpose. Orbital elements of nine solutions (two solutions for WDS 17375+2419) are listed in Table 1 where the first two columns give the WDS designation and discoverer designation of the binary, while the remaining columns list respectively the  period (in years), time of periastron passage (in years), eccentricity, semimajor axis (in arcseconds), inclination (in degrees), node (in degrees), and longitude of periastron (in degrees), together with their standard errors. Apparent orbits are graphically represented in Fig. 1. All orbits were previously announced in the IAU Commission 26 Information Circular No. 182 (Docobo et al. 2015). In order to numerically assess the quality of the obtained solutions, we report in the self-descriptive Table 2 (columns 2-6), the weighted root mean squares (RMS) and arithmetic means (AM) of (O$-$C) residuals in $\theta$ and $\rho$ for both newly calculated and previously known orbits, using the observation weighting rules described in Docobo \& Ling (2003). Biannual ephemerides for short period ($<$ 50 yr) orbits are included in Table 3, from 2016.0 to 2020.0. In Table 4 we give the yearly ephemerides for long period orbits ($>$ 50 yr) from 2016.0 to 2025.0.


\section{Systemic masses}

Some astrophysical parameters along with total mass and parallaxes are also presented in Table 2 where the first column lists the WDS designation and the apparent magnitude and spectral type of each component. Individual spectral types were assigned following the method of Edwards (1976) with the magnitude difference taken from the INT4 Catalog. In the same column, we include the Hipparcos parallax, $\pi_{Hip}$ (van Leeuwen 2007). The dynamical parallax, $\pi_{dyn}$ (Baize \& Romani, 1946; Heintz 1978a, p. 62), obtained from an improved orbit using the updated M-L and M-Sp. standard calibrations (Scmidt-Kaler 1982; Gray 2005; Docobo \& Andrade 2013), is shown in column 7. The last two columns represent systemic mass calculated on the basis of the Hipparcos and the calculated dynamical parallaxes. Column 8 gives the total mass using the Hipparcos parallax, and column 9, using the dynamical parallax.

We used the orbit grading method described in the Sixth Catalog of Orbits of Visual Binary Stars (hereinafter ORB6; Hartkopf et al. 2001b) where Grade 1 corresponds to the definitive orbit and Grade 5 is assigned to a preliminary orbit. In three cases, the grading has improved by one unit.

\section{Comments on individual stars}

\subsection{WDS 02366+1227 (=MCA 7)}

The first almost circular (P$=$3.87 yr; a$=$0.077; e$=$0.037) orbit was reported by Balega \& Balega (1988). It has been revised by Mason (1997) who, twisting some position angles by 180$^{\circ}$, transformed this solution into a highly elliptical and shorter period solution (P$=$1.92 yr; a$=$0.119; e$=$0.884), with better observational residuals (mainly in $\theta$) but leading to an unrealistic mass sum, $\mathcal{M}$=$19.1\mathcal{M}_{\odot}$, using $\pi_{Hip}$ and a small mass, 2.04, using the $\pi_{dyn}$. In view of these values, Balega's solution should have been closer to the correct one but was also demonstrating some unacceptable observational residuals. Therefore, we calculated an  orbit with a slightly smaller period (P$=$3.80 yr; a$=$0.077; e$=$0.017) which provided better observational residuals in comparison with both previous solutions (see Table 2) and with a total mass similar to that reported by Balega. Although the Grade of this orbit changed from 2 to 1, that is to say, almost definitive, we concluded that the parallax which gives more realistic masses for these F7V stars is, in this case, the dynamical parallax. The orbit is shown in Figure 1.

\subsection{WDS 02434-6643 (=FIN 333)}

We propose an orbit with an inclination of 90$^{\circ}$ and a much smaller excentricity than previous solutions. Since the magnitude difference, dm$=$0.4 mag, is obtained from speckle measurements, we do not suggest the use of dm=1.7 mag appearing in the ORB6 Catalog (Harkopf et al. 2001b). Taking into account the combined magnitude, 6.26 mag, taken from SIMBAD, we adopted 6.83 mag and 7.23 mag for the components. Applying the $\pi_{Hip}$=$18.40\pm0.43$ mas, we readily obtain $\mathcal{M}$=$3.0\mathcal{M}_{\odot}$ which is very compatible with F4+F7 dwarfs. Curiously, the masses obtained are the same as that of the improved orbit of S\"oderhjelm (1999). Grade 3 remains unchanged. There would be an eclipse in this system as long as the inclination is between 89$^{\circ}$.7 and 90$^{\circ}$3. In the most favourable case (90$^{\circ}$), the eclipse would last 7 days, from 2040.3315 to 2040.3505. Figure 2 corresponds to this orbit.

\subsection{WDS 03244-1539 (=A 2902)}

Two previous orbits of Muller (1955) and Starikova (1978) are rather similar and suggest an orbital period of 25 yr (a=0.15; e=0.19) yielding $\pi_{dyn}$ in order of 12.6 mas which is rather different from that of the Hipparcos (20.27 $\pm$0.73mas). The latest speckle measurement of Tokovinin et al. (2014) led to very large rms in both $\theta$ and $\rho$ (35 deg. and 70 mas, respectively) indicating the need for orbit improvement. Despite a relatively small set of observations and a large gap in successful observations between 1962 and 2013, we  were able to improve previous solutions and suggest a rather new elliptical orbit (e = 0.507) with a much shorter period of 11.35 yr. Not only the latest measurement is perfectly well located on the new orbital path (see Figure 3) but the global rms values are significantly improved as well (see Table 2). However, the $\pi_{dyn}$=$28.89$ mas is still far from that of Hipparcos. The total mass using the dynamical parallaxes from both previous solutions is $2.4\mathcal{M}_{\odot}$, whereas we obtained $1.6\mathcal{M}_{\odot}$ which seems to be more reasonable for a pair of early G dwarfs. The Grade improved from 4 to 3.

\subsection{WDS 08507+1800 (=A 2473)}

The reprocessed $\pi_{Hip}$=$2.77\pm1.03$ mas is clearly erroneous because it leads to absolutely unrealistic values for the mass sum of 372 $\mathcal{M}_{\odot}$ and an absolute magnitude of -0.3 mag for the G5 type. This fact indicated that we should study this system. A previous orbit (Hartkopf \& Mason 2000; P=116.7; a=0.476) led to $\pi_{dyn}$=$13.82$ mas (72.4 pc) and total mass of 2.6 $\mathcal{M}_{\odot}$ which is too large for a pair of G5 stars. Apart from this, an absolute magnitude, M=+3.3, for the main component  is obtained which is not typical for such a star. We tried to improve this orbit and got a better solution whit a similar period but a larger semiaxis (P=113.39, a=0.979, Figure 4) which yields $\pi_{dyn}$=$34.37$ (distance 29.1 pc) and typical values for both mass, $1.8\mathcal{M}_{\odot}$, and absolute magnitude of the main component (+5.5 mag). Therefore, we suggest the use of 29.1 pc as a realistic distance estimate for this star. This example shows that the Hipparcos parallax should be used with caution for distant stars ($\pi<$5 mas). The Grade remains 3.

\subsection{WDS 09128-6055 (=HDO 207)}

The new orbit gives $\pi_{dyn}$=$5.82$ mas and a mass sum of 5.2 $\mathcal{M}_{\odot}$ which is quite reasonable for a couple of late B type stars (the magnitude difference is 0.2 mag). Previous orbits of Mason \& Hartkopf (2011) and Heintz (1996) reported much shorter periods of 78.5 yr (a=0.343) and 71.3 yr (a=0.265), respectively, and dynamical masses of 34$\mathcal{M}_{\odot}$ and 19$\mathcal{M}_{\odot}$ which are too large. This example demonstrates that by only improving the orbit, one can obtain a reliable mass with a parallax value that is very close to that of Hipparcos in spite of being a rather distant pair. The Grade remains 3. The orbit is drawn in Figure 5.

\subsection{WDS 11532-1540 (=A 2579)}

The orbit by Hartkopf \& Mason (2010) with P=145 yr, a=0.403, and e=0.287 gives a mass sum $2.4\mathcal{M}_{\odot}$ but places this star at $\pi_{dyn}$=$10.87$ mas which is rather different than $\pi_{Hip}$ (3.74 $\pm$ 1.05 mas). We have calculated a 236.3 yr orbit (Figure 6) that is almost two times greater with a rather elliptical (e=0.703) solution and a somewhat smaller semiaxis (a=0.321) that give a total mass of 3.1 $\mathcal{M}_{\odot}$ (which is more reasonable for two A9+F1 stars) as well as $\pi_{dyn}$=$5.77$ mas that is rather close to that of $\pi_{Hip}$. The Grade remains 3

\subsection{WDS 17375+2419 (=CHR 63)}
There are two possible solutions: a short period, eccentric orbit (P$_{1}$=10.42 yr; e$_{1}$=0.753) and a long period, almost circular orbit (P$_{2}$=20,91 yr; e$_{2}$=0.027) that improves the earlier orbits and leads to dynamical parallaxes of 7.6 mas and 17.4 mas, respectively. The total mass obtained in solution II (3.06$\mathcal{M}_{\odot}$) seems more reasonable for a pair of A1+A8 dwarfs than that of solution I (4.65$\mathcal{M}_{\odot}$) and its $\pi_{dyn}$ is closer to that of Hipparcos (13.04$\pm$0.31 mas). The rms of residuals are slightly better for solution I (see Table 2). The Grade remains 3. Figures 7 (long period) and 8 (short period) correspond to this system.

\subsection{WDS 22408-0333 (=KUI 114)}

We have obtained two improved solutions corresponding to a reasonable dynamical mass value. The short 27 yr period solution improves those of Baize (1976) and S\"{o}derhjelm (1999) (both with an orbital period of 28 yr) but largely disagrees with the RV data reported by Griffin \& Heintz (1987). The long  54 yr period solution agrees well with the RV data and leads to mass sum $2.5\mathcal{M}_{\odot}$ and $\pi_{dyn}=18.84$ mas which differs from $\pi_{Hip}=28.93\pm0.77$ mas. Interestingly, the short period solution yields $2.3\mathcal{M}_{\odot}$ and $\pi_{dyn}=23.03$ mas which are somewhat closer to that of Hipparcos but have unacceptably large rms residuals. Therefore, the use of the long period improved orbit and $\pi_{dyn}=18.84$ mas as the distance estimate is likely the best choice.

Our orbit is very similar to that of Griffin \& Heintz (1987) although it improves it slightly, and can be seen in Figure 9.. The Grade changes from 3 to 2.


\section*{Acknowledgements}
This research was supported by the Spanish Ministry of Economy and Competitiveness under the Project AYA2011-26429 and the IEMath-Galicia Network (R2014/002, FEDER-Xunta of Galicia). The work was partially supported by the Program of Fundamental Researches of the Presidium of RAS (P-41), by the Russian Foundation for Basic Research (projects 15-02-04053 and 16-07-01162), and by the Program of the support of leading scientific schools of RF (3620.2014.2). This research has made use of the Washington Double Star Catalogs maintained at the U.S. Naval Observatory (Mason et al. 2001) and the SIDONIe Database maintained at the Nice-C$\hat{o}$te d'Azur Observatory, France (Le Contel et al. 2001).


\begin{table*}
\begin{minipage}{\textwidth}
  \caption{Orbital elements}
\begin{tabular}{clrrrrrrr}
\hline
\multicolumn{2}{c}{Star}  & \multicolumn{7}{c}{Orbital elements}\\
\cline{3-9}\\
WDS  & \multicolumn{1}{c}{Name} & \multicolumn{1}{c}{P (yr)} & \multicolumn{1}{c}{T} &  \multicolumn{1}{c}{e} & \multicolumn{1}{c}{a ($^{\prime\prime}$)}  & \multicolumn{1}{c}{i ($^{\circ}$)} & \multicolumn{1}{c}{$\Omega$ ($^{\circ}$)} & \multicolumn{1}{c}{$\omega$ ($^{\circ}$)}\\
02366+1227   & MCA 7 &3.80$\pm$0.10   & 2010.28$\pm$0.15  & 0.017$\pm$0.002  & 0.077$\pm$0.001 & 112.7$\pm$0.5  & 145.0$\pm$0.5  & 3.7$\pm$15.0\\ 
02434$-$6643   & FIN 333& 83.73$\pm$2.50 & 1998.06$\pm$0.50  & 0.423$\pm$0.005  & 0.509$\pm$0.003 & 90.0$\pm$0.5 & 34.1$\pm$1.0  & 269.2$\pm$2.0\\
03244$-$1539  & A 2902& 11.35$\pm$0.26& 2013.74$\pm$0.05  & 0.507$\pm$0.015  & 0.172$\pm$0.003 & 71.3$\pm$1.0   & 16.2$\pm$0.5  & 283.0$\pm$1.0\\
08507+1800  & A 2473&113.4$\pm$1.0    & 2018.01$\pm$0.50  & 0.972$\pm$0.010  & 0.979$\pm$0.040 &  80.2$\pm$1.0  &115.0$\pm$1.0  &  89.0$\pm$1.0\\ 
09128$-$6055 & HDO 207 &400.$\pm$30. & 1960.65$\pm$1.65  & 0.538$\pm$0.050  & 0.548$\pm$0.007 &  76.4$\pm$1.5  & 43.7$\pm$1.5 & 149.0$\pm$5.0\\ 
11532$-$1540 & A 2579 &236.3$\pm$20.0   & 1897.68$\pm$1.65  & 0.287$\pm$0.050  & 0.321$\pm$0.005 & 147.9$\pm$3.0  & 17.8$\pm$5.0  & 160.6$\pm$15.0\\ 
17375+2419   &CHR 63 I& 20.92$\pm$0.20 & 1992.28$\pm$0.30  & 0.027$\pm$0.025  & 0.096$\pm$0.003 & 114.8$\pm$2.0  & 62.9$\pm$2.0 &  97.9$\pm$15.0\\ 
17375+2419   &CHR 63 II& 10.42$\pm$0.15 & 1995.51$\pm$0.20  & 0.753$\pm$0.010  & 0.127$\pm$0.003 & 100.6$\pm$1.5  & 72.4$\pm$2.0  & 259.8$\pm$1.5\\ 
22408$-$0333 & KUI 114 & 54.57$\pm$0.30  & 1969.38$\pm$1.50  & 0.005$\pm$0.003  & 0.367$\pm$0.003 &  87.5$\pm$0.5  &128.9$\pm$0.5  & 12.9$\pm$20.0\\ 
\hline
\end{tabular}
\end{minipage}
\end{table*}

\begin{table*}
\begin{minipage}{\textwidth}
  \caption{Quality indicators}
\begin{tabular}{l|l|rcrrrrr}
\hline

\multicolumn{1}{l|}{WDS}& \multicolumn{1}{l|}{Authors}&
\multicolumn{2}{c}{\quad RMS}& \multicolumn{2}{c}{\quad\quad MA}& \multicolumn{1}{c}{$\pi_{dyn} \pm \sigma $} & \multicolumn{2}{c}{Total mass}
\\
\multicolumn{1}{l|}{Magnitudes}&\multicolumn{1}{l|}{} & & & & & & & \\
\multicolumn{1}{l|}{Sp. Types} & 
\multicolumn{1}{l|}{} & \multicolumn{1}{c}{\quad $\Delta \theta$($
^\circ$)}& \multicolumn{1}{c}{$\Delta \rho$ ($^{\prime\prime}$) }&
\multicolumn{1}{c}{\quad\quad $\Delta \theta$($ ^\circ$)}&
\multicolumn{1}{c}{$\Delta \rho$($^{\prime\prime}$) } & \multicolumn{1}{c}{(mas)}&
\multicolumn{1}{c}{Hip}&
\multicolumn{1}{c}{Dyn}\\
\multicolumn{1}{l|}{$\pi_{Hip}\pm \sigma $ (mas)} & \multicolumn{1}{l|}{} & & & & & & & \\
\hline\\

02366+1227 &Docobo et al. (2015)& 8.1&0.007&0.0& 0.001 & 21.23$\pm$0.55 & 1.33 & 3.36$\pm$0.04 \\ %
5.68-5.78 &Mason (1997) & 15.9&0.012&$-$6.9& $-$0.004 & 60.68 & 19.08 & 2.04 \\
F7V-F7V & Balega \& Balega (1988) &26.1&0.012&$-$14.8& 0.000 & 20.82 & 1.28 & 3.38 \\
28.79$\pm$0.43 & & & & & & & & \\
\hline
02434$-$6643 &Docobo et al. (2015) & 1.7&0.030&0.0& 0.001 & 19.18$\pm$0.47 & 3.02 & 2.67$\pm$0.03 \\ %
6.83-7.23 & Mason \& Hartkopf (2011) & 2.2&0.033&$-$0.7& $-$0.001 & 35.89 & 14.55 & 1.96 \\
F4V-F7V & S\"oderhjelm (1999) & 3.3&0.032&1.3& $-$0.014 & 19.27 & 3.02 & 2.63 \\
18.40$\pm$0.43 & Heintz (1978b) & 2.2&0.196&$-$1.3& 0.050 & 22.14 & 4.29 & 2.46 \\
\hline
03244$-$1539  &Docobo et al. (2015)& 3.0&0.021&1.4& 0.011 & 28.89$\pm$0.79 & 4.74 & 1.64$\pm$0.02 \\ %
8.40-8.40 & Starikova (1978) & 34.5 & 0.064 & 27.6 & $-$0.057 & 12.85 & 0.61 & 2.40 \\
G1V-G1V & Muller (1955) & 32.9 & 0.063 & 25.7 & $-$0.055 & 12.93 & 0.62 & 2.40 \\
20.27$\pm$0.73 & & & & & & & & \\
\hline
08507+1800 &Docobo et al. (2015)& 2.8&0.014&$-$0.2& $-$0.002 & 34.37$\pm$1.68 & 3433.07 & 1.80$\pm$0.04 \\ %
7.56-7.66 & Hartkopf \& Mason (2000) & 3.1&0.019&$-$0.7& 0.007 & 14.27 & 372.60 & 2.62 \\
G5V-G5V & Scardia (1984) & 3.2 & 0.032 & $-$1.0 & $-$0.018 & 5.18 & 28.74 & 4.40 \\
2.77$\pm$1.03 & & & & & & & & \\
\hline
09128$-$6055 &Docobo et al. (2015)& 5.1&0.016&1.0& $-$0.001 & 5.82$\pm$0.36 & 5.35 & 5.22$\pm$0.14 \\ %
6.97-7.27 & Mason \& Hartkopf (2011) & 10.3&0.020&2.5& 0.007 & 12.10 & 34.09 & 3.69 \\
B9V-B9V & Heintz (1996) & 18.8&0.026&$-$9.8& 0.008 & 9.61 & 19.06 & 4.12 \\
5.77$\pm$0.48 & & & & & & & & \\
\hline
11532$-$1540  &Docobo et al. (2015)& 3.5&0.014&0.3& $-$0.002 & 5.77$\pm$0.40 & 11.32 & 3.08$\pm$0.10 \\ %
8.61-9.29 & Hartkopf \& Mason (2010) & 3.5 & 0.012 & 0.4 & $-$0.002 & 11.11 & 59.26 & 2.25 \\
A9V-F1V & Baize (1981) & 4.1 & 0.039 & 2.0 & 0.030 & 6.49 & 15.24 & 2.91 \\
3.74$\pm$1.05 & & & & & & & & \\
\hline
17375+2419 I & Docobo et al. (2015)& 2.1&0.004&0.0& 0.000 & 7.58$\pm$0.29 & 0.91 & 4.65$\pm$0.08 \\ %
5.90-7.30 & Hartkopf et al. (2000) & 5.5 & 0.007 & 3.2 & 0.002 & 22.18 & 13.76 & 2.80 \\
A1V-A8V & Olevic \& Jovanovic (1998) & 8.3 & 0.010 & 5.4 & 0.006 & 7.73 & 0.96 & 4.60 \\
13.04$\pm$0.31 & & & & & & & & \\
\hline
17375+2419 II &Docobo et al. (2015)& 1.6&0.011&0.0& 0.002 & 17.40$\pm$0.55 & 8.51 & 3.06$\pm$0.04 \\ %
5.90-7.30 & Hartkopf et al. (2000) & 5.5 & 0.007 & 3.2 & 0.002 & 22.18 & 13.76 & 2.80 \\
A1V-A8V & Olevic \& Jovanovic (1998) & 8.3 & 0.010 & 5.4 & 0.006 & 7.73 & 0.96 & 4.60 \\
13.04$\pm$0.31 & & & & & & & & \\
\hline
22408$-$0333 &Docobo et al. (2015)& 3.0&0.016&$-$0.2& 0.000 & 18.47$\pm$0.20 & 0.69 & 2.48$\pm$0.01\\        %
6.52-8.63 & S\"oderhjelm (1999) & 5.7&0.023&0.3& $-$0.009 & 19.98 & 0.80 & 2.42 \\
F6V-G9V & Griffin \& Heintz (1987) & 3.1&0.018&$-$0.6& 0.001 & 18.51 & 0.66 & 2.50 \\
28.93$\pm$0.77 & & & & & & & & \\
\hline
\end{tabular}
\end{minipage}
\end{table*}

\begin{table*}
\begin{minipage}{126mm}
\caption{Ephemerides for short period orbits}
\begin{tabular}{r|cc|cc|cc|cc}
\hline
Epoch & \multicolumn{2}{c}{MCA 7} & \multicolumn{2}{c}{A 2902} & \multicolumn{2}{c}{CHR 63 I} & \multicolumn{2}{c}{CHR 63 II} \\
 \cline{2-9}\\
 & $\theta$ & $\rho$ & $\theta$& $\rho$ & $\theta$ & $\rho$ & $\theta$ & $\rho$ \\
\hline\\
2016.0 &$\;\;$17.8  &0.068 &$\;\;$38.4 &0.125 &257.3 &0.083 &253.8 &0.053\\
2016.5 &358.7 &0.076 &$\;\;$47.5  &0.113 &252.8 &0.089 &$\;\;$84.8 &0.024\\
2017.0 &330.3 &0.046 &$\;\;$59.0 &0.100 &248.8 &0.093 &$\;\;$69.9 &0.077\\
2017.5 &233.9 &0.036 &$\;\;$73.8 &0.088 &245.0 &0.095 &$\;\;$65.5 &0.096\\
2018.0 &193.1 &0.071 &$\;\;$92.1 &0.081 &241.4 &0.096 &$\;\;$62.1 &0.103\\
2018.5 &173.3 &0.070 & 111.8&0.082 &237.7 &0.094 &$\;\;$59.0 &0.104\\
2019.0 &132.2 &0.036 & 129.8&0.089 &233.8 &0.091 &$\;\;$55.8 &0.101\\
2019.5 &$\;\;$38.2 &0.046 & 144.1&0.102 &229.6 &0.087 &$\;\;$52.3 &0.095\\
2020.0 &$\;\;\;\;$9.5 &0.076 & 154.9&0.117 &224.9 &0.081 &$\;\;$48.2 &0.087\\
\hline
\multicolumn{9}{l}{\footnotesize{Position angles and separations are given in degrees and arcseconds,}}\\
\multicolumn{9}{l}{\footnotesize{respectively}}
\end{tabular}
\end{minipage}
\end{table*}


\begin{table*}
\begin{minipage}{126mm}
\caption{Ephemerides for long period orbits}
\begin{tabular}{c|cc|cc|cc|cc|cc}
\hline
Epoch & \multicolumn{2}{c}{FIN 333} & \multicolumn{2}{c}{A 2473} & \multicolumn{2}{c}{HDO 207} & \multicolumn{2}{c}{A 2579} & \multicolumn{2}{c}{KUI 114}\\
 \cline{2-11}\\
 & $\theta$ & $\rho$ & $\theta$& $\rho$ & $\theta$ & $\rho$ & $\theta$ & $\rho$ & $\theta$& $\rho$\\
\hline\\
2016.0 &34.1 &0.457 &$\;\;$98.2 &0.171 &285.7 &0.132 &34.3 &0.407  &126.8 &0.280 \\
2017.0 &34.1 &0.451 &103.4 &0.134 &288.8 &0.130 &33.5 &0.407  &127.3 &0.306 \\
2018.0 &34.1 &0.442 &157.6 &0.007 &292.1 &0.128 &32.7 &0.408  &127.6 &0.327 \\
2019.0 &34.1 &0.432 &305.7 &0.138 &295.4 &0.127 &32.0 &0.408  &128.0 &0.343 \\
2020.0 &34.1 &0.420 &310.5 &0.180 &298.8 &0.126 &31.2 &0.408  &128.3 &0.355 \\
2021.0 &34.1 &0.407 &313.7 &0.207 &302.2 &0.125 &30.5  &0.409  &128.6 &0.363 \\
2022.0 &34.1 &0.393 &316.3 &0.227 &305.6 &0.125 &29.7 &0.409  &128.9 &0.365 \\
2023.0 &34.1 &0.377 &318.5 &0.242 &309.0 &0.126 &28.9 &0.409  &129.2 &0.363 \\
2024.0 &34.1 &0.360 &320.5 &0.255 &312.4 &0.126 &28.2 &0.409  &129.5 &0.355 \\
2025.0 &34.1 &0.343 &322.3 &0.266 &315.7 &0.127 &27.4 &0.409  &129.8 &0.343 \\
\hline
\multicolumn{11}{l}{\footnotesize{Position angles and separations are given in degrees and arcseconds, respectively}}
\end{tabular}
\end{minipage}
\end{table*}

\begin{figure*}
\includegraphics[width=0.5\textwidth]{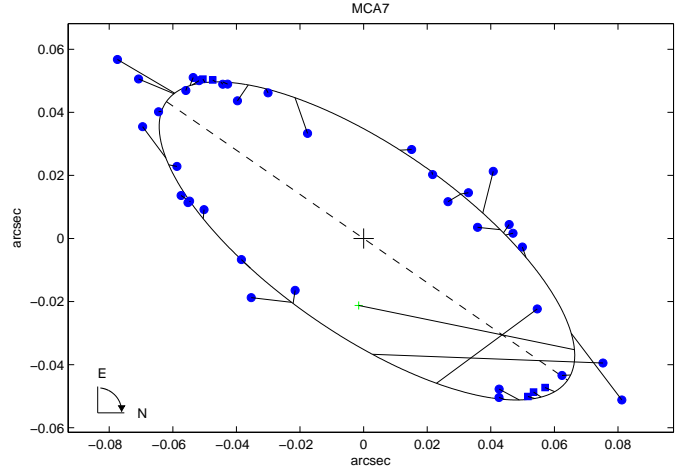}
\caption{MCA 7. The plus signs represent visual data, filled circles correspond to measurements made with speckle and other single aperture interferometry techniques, and squares stand for multiple aperture techniques.}
\label{mca7fig}
\end{figure*}

\begin{figure*}
\includegraphics[width=0.5\textwidth]{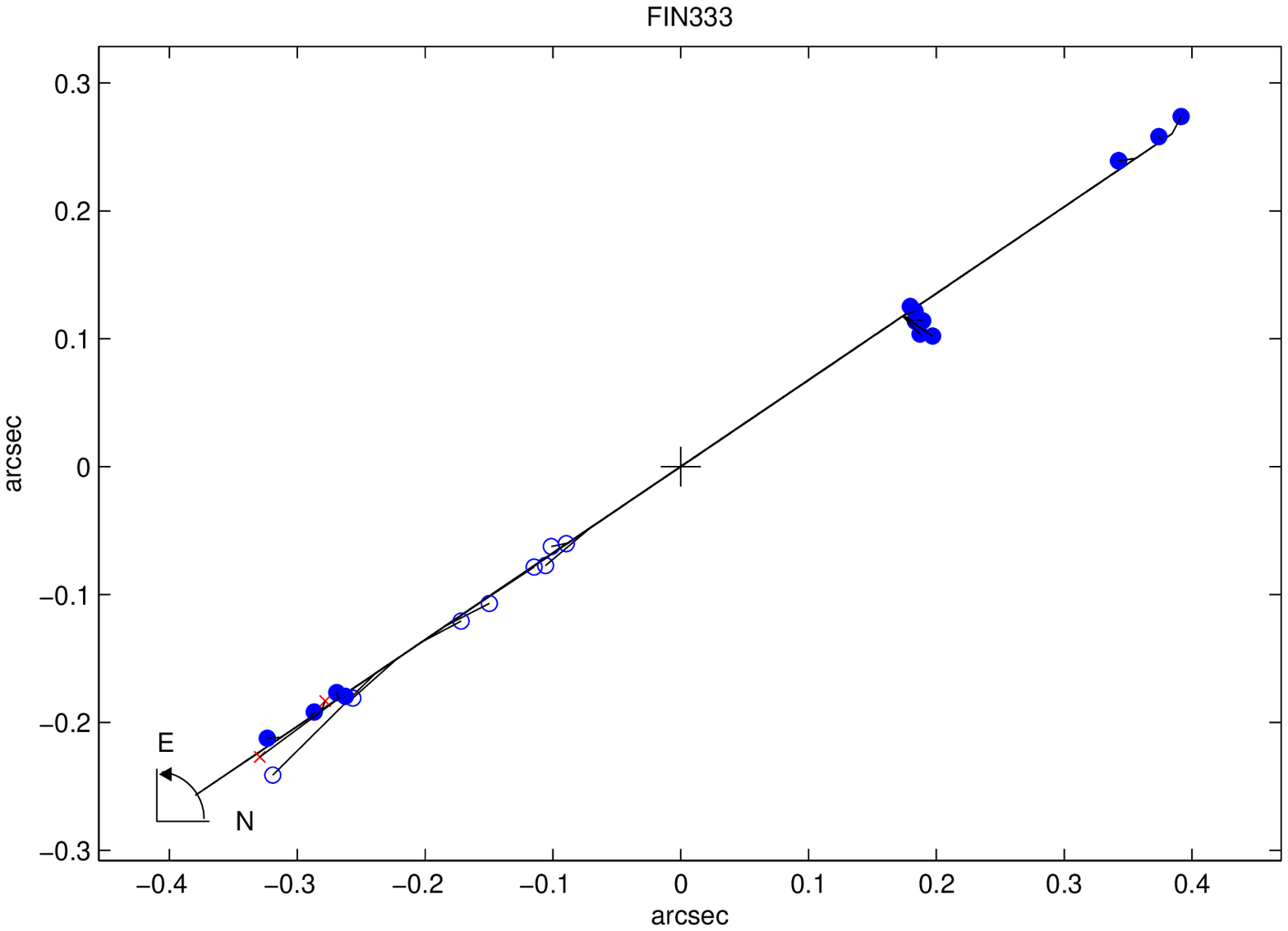}
\caption{FIN 333. The plus signs represent visual data, crosses are Hipparcos and Tycho measurements, open circles correspond to measurements made with eyepiece interferometry, and filled circles indicate speckle and other single aperture interferometry techniques.}
\label{fin333fig}
\end{figure*}

\begin{figure*}
\includegraphics[width=0.5\textwidth]{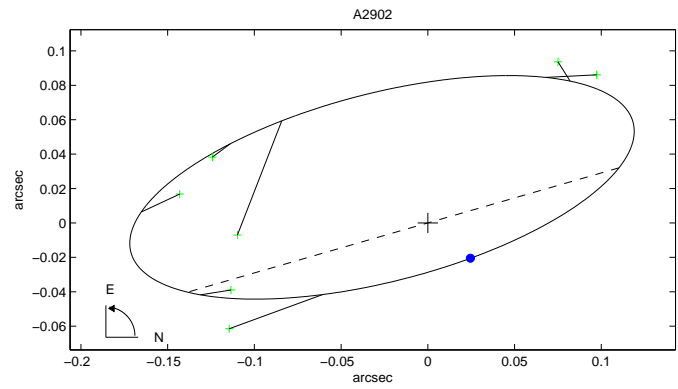}
\caption{A 2902. The plus signs represent visual data, and filled circles correspond to measurements made with speckle and other single aperture interferometry techniques.}
\label{ads2524fig}
\end{figure*}

\begin{figure*}
\includegraphics[width=0.5\textwidth]{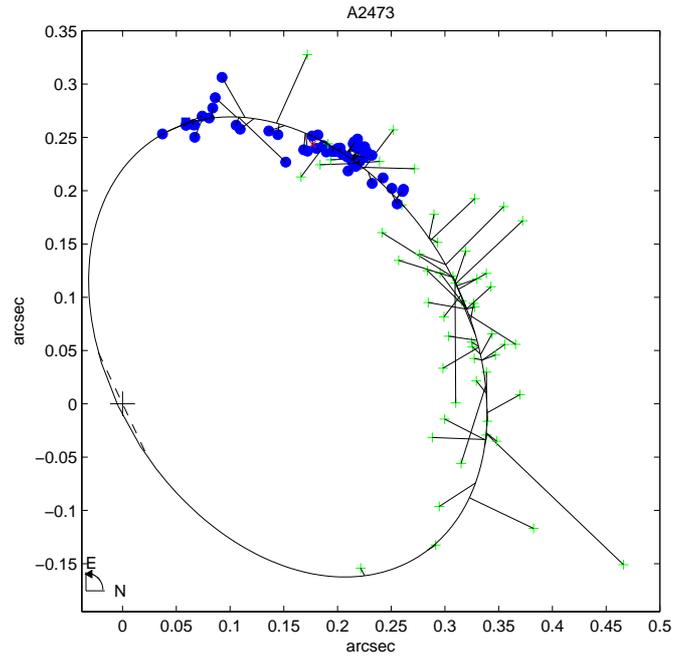}
\caption{A 2473. The plus signs represent visual data, crosses are Hipparcos and Tycho measurements, filled circles correspond to measurements made with speckle and other single aperture interferometry techniques, and filled squares indicate multi-aperture interferometry.}
\label{ads7039fig}
\end{figure*}

\begin{figure*}
\includegraphics[width=0.5\textwidth]{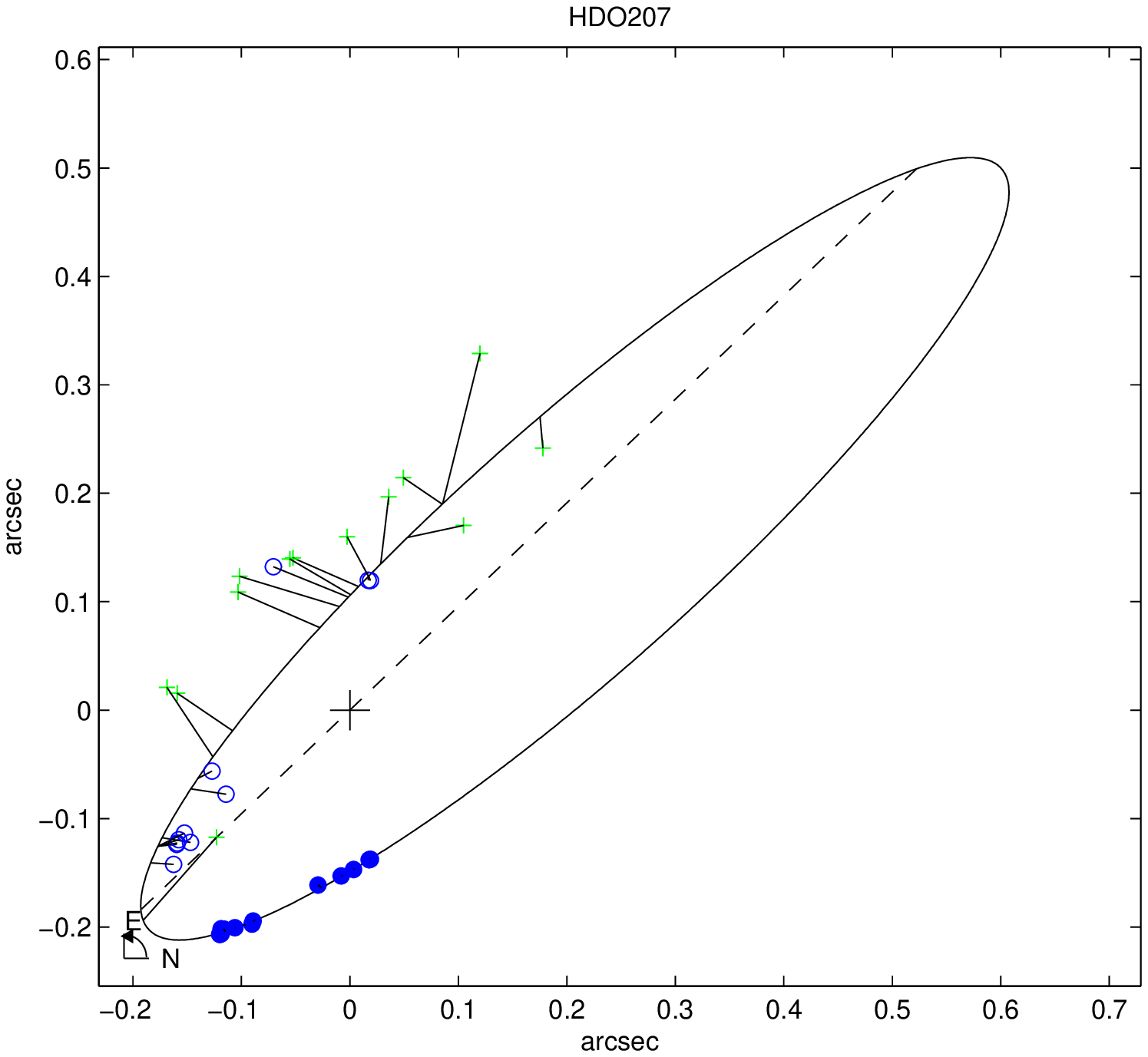}
\caption{HDO 207. The plus signs represent visual data, crosses are Hipparcos and Tycho measurements, open circles correspond to measurements made with eyepiece interferometry, and filled circles indicate speckle and other single aperture interferometry techniques}
\label{hdo207fig}
\end{figure*}

\begin{figure*}
\includegraphics[width=0.5\textwidth]{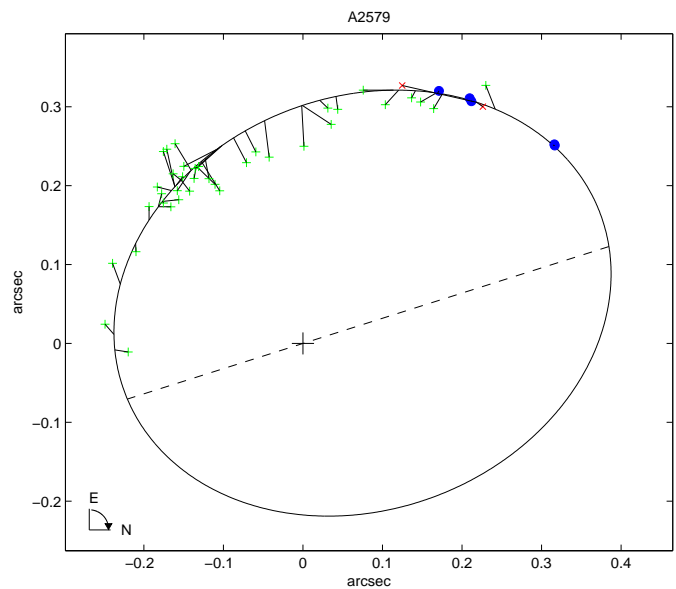}
\caption{A2579. The plus signs represent visual data, crosses are Hipparcos and Tycho measurements, and filled circles correspond to measurements made with speckle and other single aperture interferometry techniques.}
\label{ads8332fig}
\end{figure*}

\begin{figure*}
\includegraphics[width=0.35\textwidth]{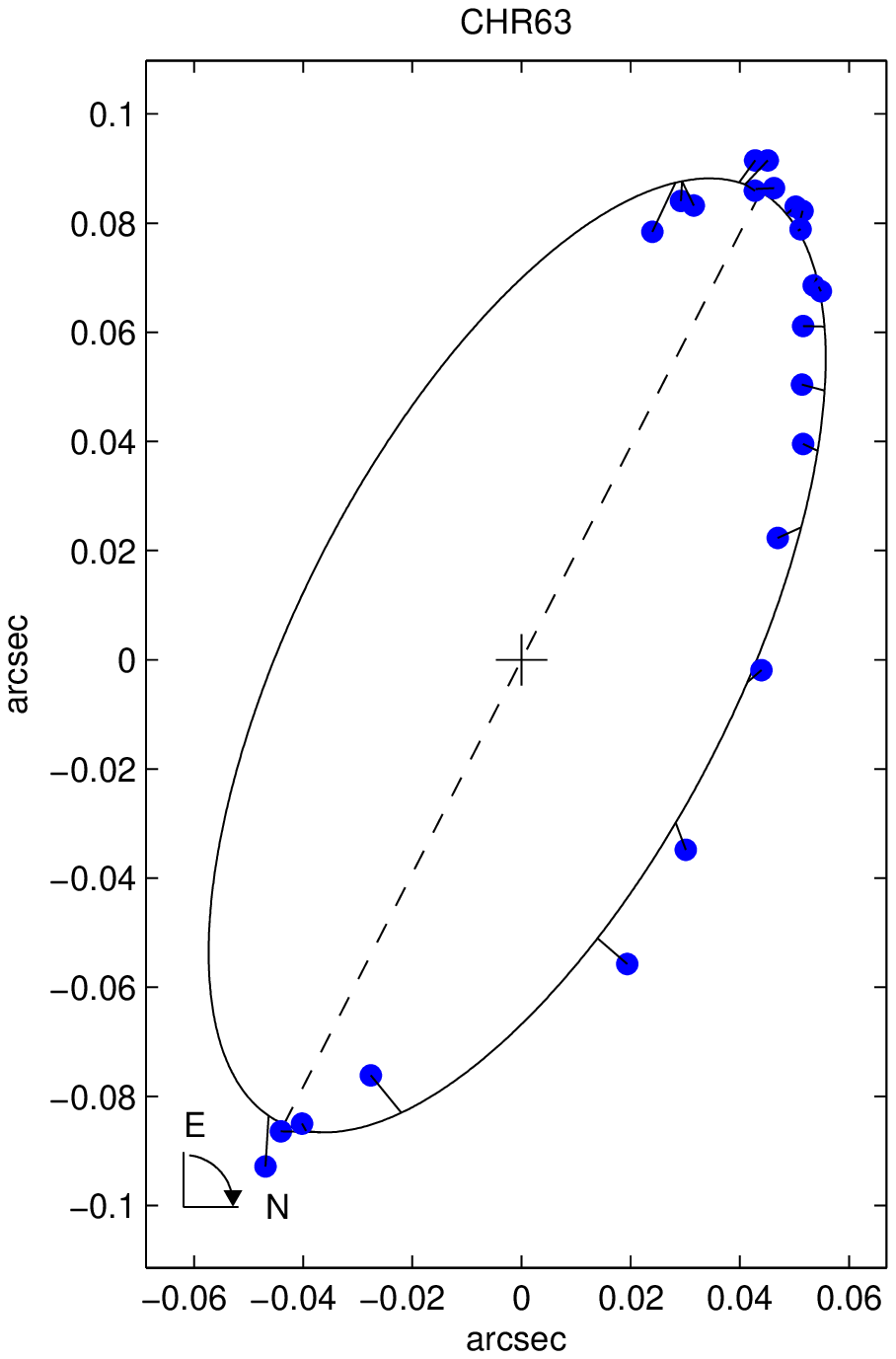}
\caption{CHR 63, long period solution. The filled circles indicate measurements made with speckle and other single aperture interferometry techniques}
\label{chr63lfig}
\end{figure*}

\begin{figure*}
\includegraphics[width=0.35\textwidth]{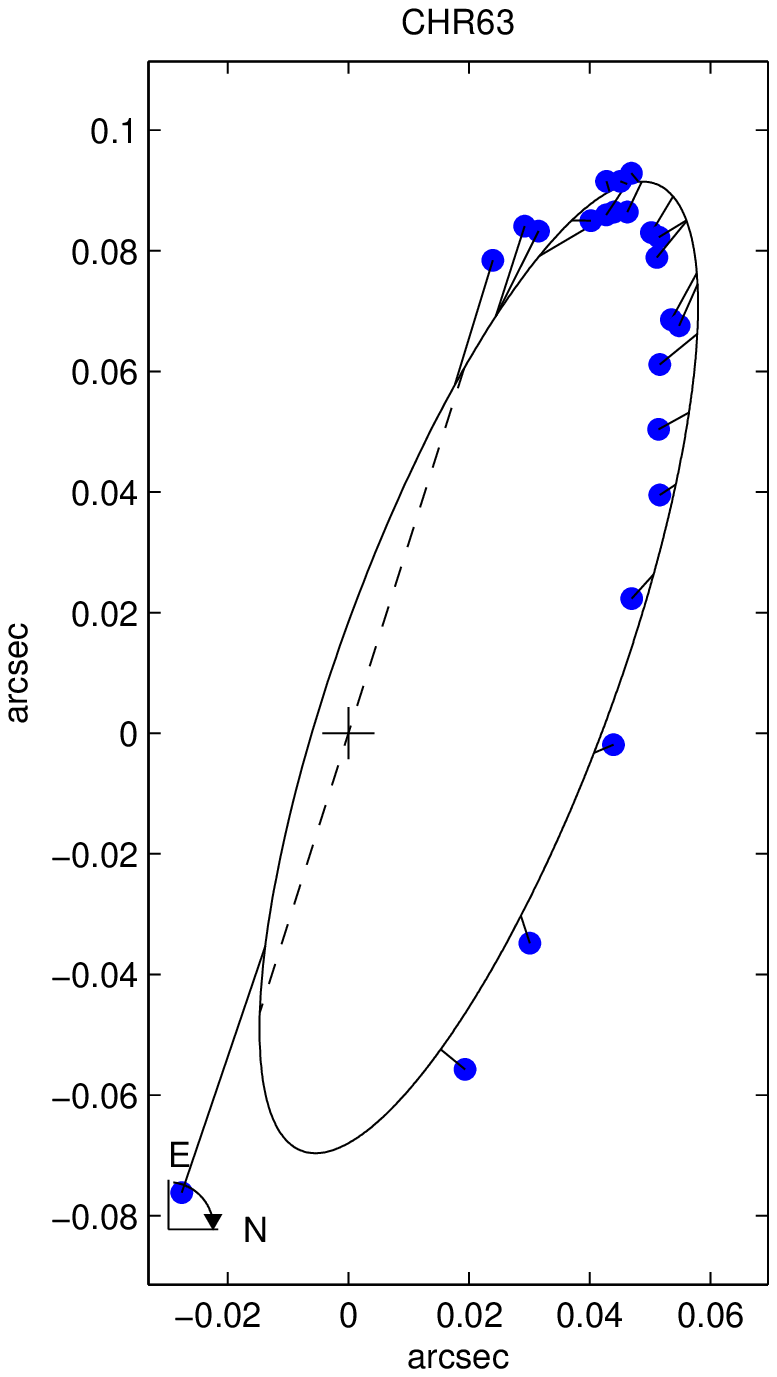}
\caption{CHR 63, short period solution. The filled circles indicate measurements made with speckle and other single aperture interferometry techniques}
\label{chr63cfig}
\end{figure*}

\begin{figure*}
\includegraphics[width=0.35\textwidth]{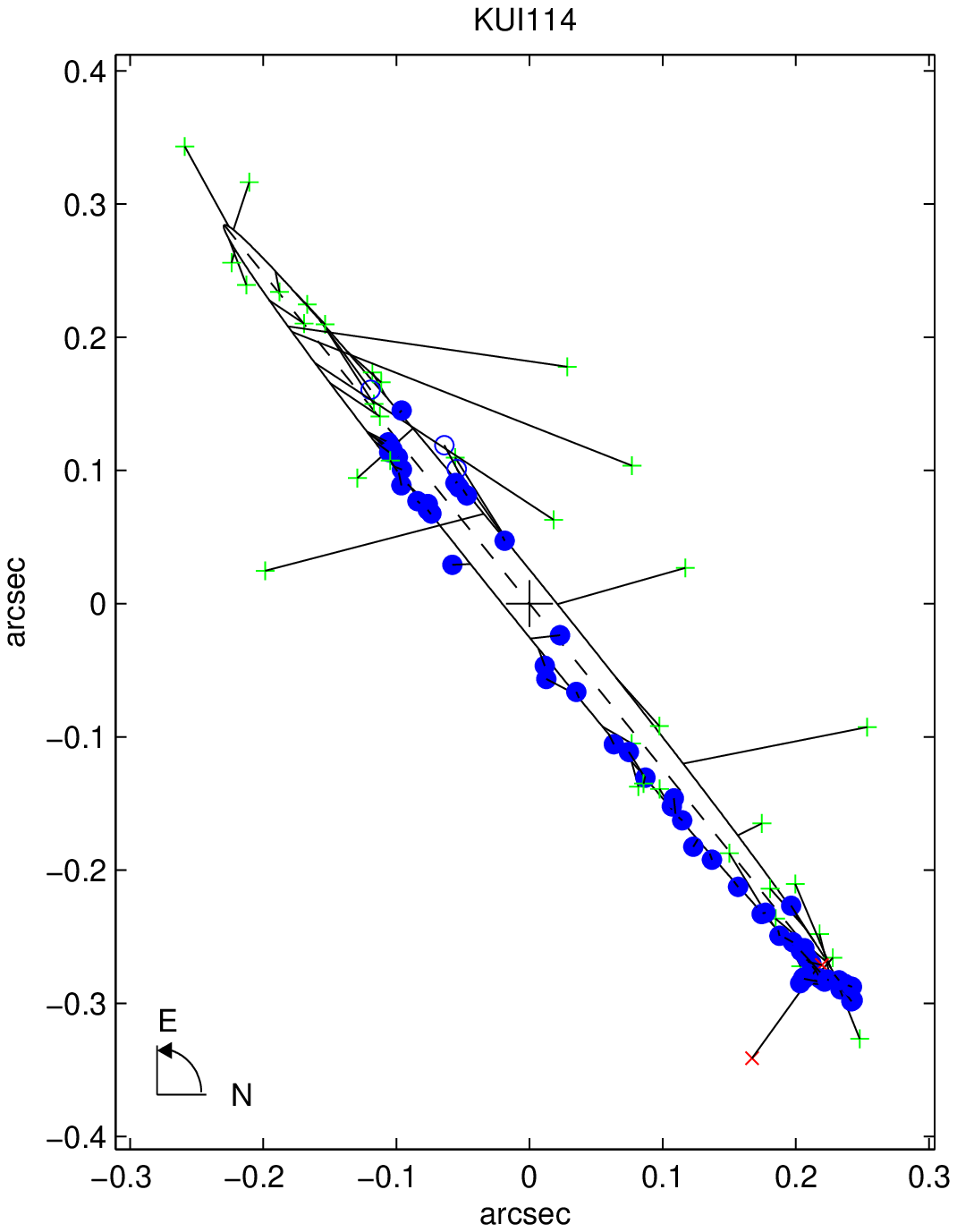}
\caption{KUI 114. The plus signs represent visual data, crosses are Hipparcos and Tycho measurements, open circles correspond to measurements made with eyepiece interferometry, and filled circles indicate speckle and other single aperture interferometry techniques}
\label{kui114fig}
\end{figure*}
\bsp
\label{lastpage}

\begin{thebibliography}{}
\bibitem[\protect\citeauthoryear{Baize}{1946}]{BR} Baize, P., \& Romani, L. 1946, Ann. d'Astrophys. 9, 13
\bibitem[\protect\citeauthoryear{Baize}{1976}]{Bai1} Baize, P., 1976, A\&AS, 26, 177
\bibitem[\protect\citeauthoryear{Baize}{1981}]{Bai2} Baize, P., 1981, A\&AS, 44, 202                                                     
\bibitem[\protect\citeauthoryear{Balega}{1988}]{Bal} Balega, Y., \& Balega I. 1998, SvAL, 14, 393                                            
\bibitem[\protect\citeauthoryear{Docobo1}{1985}]{Doc1} Docobo, J. A. 1985, Celest. Mech., 36, 143                              
\bibitem[\protect\citeauthoryear{Docobo2}{2011}]{Doc2} Docobo, J. A. 2012, in Arenou, F. \& Hestroffer D., eds., Proc. Workshop "Orbital Couples: Pas de Deux in the Solar System and the Milky Way". Paris Observatory, Paris,  p. 119
\bibitem[\protect\citeauthoryear{Docobo3}{2013}]{Doc3} Docobo, J. A., \& Andrade, M. 2013, MNRAS, 428, 321   
\bibitem[\protect\citeauthoryear{Docobo4}{2003}]{Doc4} Docobo, J. A., \& Ling, J. F.  2003, A\&A, 409, 989              
\bibitem[\protect\citeauthoryear{Docobo5}{2015}]{Doc5} Docobo, J. A., Tamazian, V. S., Malkov, O. Yu., Campo, P. P. \& Chulkov, D. I. 2015, IAU Commission 26 Circ. 185
\bibitem[\protect\citeauthoryear{Edwards}{1976}]{Edw} Edwards, T. W. 1976, AJ, 81, 245
\bibitem[\protect\citeauthoryear{Gray}{2005}]{Gra} Gray, D. F. 2005, in The Observation and Analysis of Stellar Photospheres. Cambridge Univ. Press. Cambridge, p. 506
\bibitem[\protect\citeauthoryear{Griffin}{1987}]{Gri} Griffin, R. F., \& Heintz, W. D.  1987, JRASC, 81, 3                       
\bibitem[\protect\citeauthoryear{Hartkopf1}{2000}]{Har1} Hartkopf, W, I., \& Mason, B. D. 2000, IAU Commission 26, Circ. 142         
\bibitem[\protect\citeauthoryear{Hartkopf2}{2010}]{Har2} Hartkopf, W.I. \& Mason, B.D.  2010, IAU Commission 26, Circ. 170                    
\bibitem[\protect\citeauthoryear{Hartkopf3}{2000}]{Har3} Hartkopf, W, I., Mason, B. D., \& McAlister, H. A. et al. 2000, AJ, 119, 3084         
\bibitem[\protect\citeauthoryear{Hartkopf4}{2001}]{Har4} Hartkopf, W. I., Mason, B.D., Wycoff, G.L., \& McAlister, H.A. 2001a, AJ, 122, 3480
\bibitem[\protect\citeauthoryear{Hartkopf5}{2001}]{Har5} Hartkopf, W. I., Mason, B.D., \& Worley C. E. 2001b, AJ, 122, 3472                               
\bibitem[\protect\citeauthoryear{Heintz1}{1978}]{Hei1} Heintz, W. D. 1978a, Double Stars. Reidel, Dordrecht
\bibitem[\protect\citeauthoryear{Heintz2}{1978}]{Hei2} Heintz, W. D. 1978b, ApJS, 37, 515                                                   
\bibitem[\protect\citeauthoryear{Heintz3}{1996}]{Hei3} Heintz, W. D. 1996, AJ, 111, 412                                                      
\bibitem[\protect\citeauthoryear{Lecontel}{2001}]{Lec} Le Contel, D., Valtier, J.-C., \& Bonneau, D. 2001, A\&A, 377, 496       
\bibitem[\protect\citeauthoryear{Mason1}{1997}]{Mas1} Mason, B. D. 1997, AJ, 114, 808                                                  
\bibitem[\protect\citeauthoryear{Mason2}{2011}]{Mas2} Mason, B. D. \& Hartkopf, W.I.  2011, IAU Commission 26, Circ. 173    
\bibitem[\protect\citeauthoryear{Mason3}{2001}]{Mas3} Mason, B. D., Wycoff, G.L., Hartkopf, W.I., Douglass, G.G., \& Worley, C.E. 2001, AJ, 122, 3466   
\bibitem[\protect\citeauthoryear{Muller}{1955}]{Mul} Muller, P. 1955, JO, 38, 17                                                   
\bibitem[\protect\citeauthoryear{Olevic}{1998}]{Ole} Olevic, D.\& Jovanovic, P.  1998, IAU Commission 26, Circ. 136  
\bibitem[\protect\citeauthoryear{Scardia}{1984}]{Sca} Scardia, M. 1984, IAU Commission 26, Circ. 92  
\bibitem[\protect\citeauthoryear{Schmidt}{1982}]{SK} Schmidt-Kaler, T. 1982, in Schaifers, K. \& Voigt, H. H., eds. Landolt-Bornstein, New Series, vol. 2b, Springer, Berlin, p. 18
\bibitem[\protect\citeauthoryear{Soderhjelm}{1999}]{Sod} S\"oderhjelm, S. 1999, A\&A, 341, 121                                           
\bibitem[\protect\citeauthoryear{Starikova}{1978}]{Sta} Starikova, G. A. 1978, SvAL, 4, 52                                              
\bibitem[\protect\citeauthoryear{Tamazian}{2016}]{Tam} Tamazian, V. S., Malkov. O., Docobo, J. A., Chulkov, D. A., Campo, P. P. 2016, ApSS, 361, art. Id. 105                                          
\bibitem[\protect\citeauthoryear{Tokovinin}{2014}]{Tok} Tokovinin, A., Mason, B. D., \& Hartkopf, W. I. 2014, AJ, 148, 72               
\bibitem[\protect\citeauthoryear{Vanleeuwen}{2007}]{VL} van Leeuwen, F. 2007, A\&A, 474, 653                                            
\end{thebibliography}
\end{document}